\documentclass[prb,preprint]{revtex4}

\usepackage{amsmath}
\usepackage{graphicx}
\usepackage[latin1]{inputenc}

\begin{document}

\title{On the exact electric and magnetic fields of an electric dipole}

\author{W.J.M. Kort-Kamp}
\email{kortkamp@if.ufrj.br}
\author{C. Farina}
\email{farina@if.ufrj.br}

\affiliation{Universidade Federal do Rio de Janeiro,
Instituto de Fisica, Rio de Janeiro, RJ 21945-970}


\begin{abstract}

We derive from Jefimenko's equations \cite{JefimenkoBook} a multipole expansion 
in order to obtain the exact expressions for the electric and magnetic fields of an electric dipole with an arbitrary time dependence.
 A few  comments are also made about the usual expositions found in most common undergraduate and graduate textbooks as well as in the literature on this topic.

 \end{abstract}

\maketitle

Even in the context of electrostatic, the problem of finding analytically the electric field associated with a localized but arbitrary static charge distribution may become quite involved. Of course, simplifications occur when symmetries are present. Due to the difficulty of getting exact solutions, numerical methods and approximation theoretical methods have been developed. One of the most important examples of the latter is the so called multipole expansion method relative to a base point (for convenience, let us take the origin at the base point). Choosing the origin inside the distribution, the multipole expansion method states, essentially, that the field outside the distribution is given by a superposition of fields, each of them being interpreted as the electrostatic field of a multipole located at the origin (see, for instance, Griffiths' textbook \cite{GriffithsBook}). The first terms of the multipole expansion correspond to the fields of a monopole,  a dipole and a quadrupole, respectively. Hence, for an arbitrary distribution with a vanishing total charge the first term in the multipole expansion is given, in principle, by the field of the electric dipole of the distribution. This fact shows the important role played by the field of an electric dipole.
Recall that a polar molecule is a neutral system whose electric field is that of an electric dipole with great accuracy.
Also for (localized) charge and current distributions with arbitrary time dependence, the multipole expansion method is an extremely powerful technique to compute the electromagnetic fields of the system. Particularly, in the study of radiation fields of simple systems, like antennas among others, it is a common procedure to make a multipole expansion and to maintain only the first contributions. It is worth emphasizing that, in contrast to the case for static charge distributions, the leading contribution for the radiation fields of a system comes from the electric dipole term, since there is no monopole radiation fields due to charge conservation. In this sense, the electromagnetic fields of an electric dipole are even more relevant for radiating systems than they are for static distributions. Besides, in the Quantum Mechanics context, dipole fields are of fundamental importance in the study of intermolecular forces.  For instance, for two neutral but polarizable atoms (or two unpolar molecules), the interatomic (intermolecular) forces can be described by the fluctuating dipole method (see Milonni's book \cite{MilonniBook} and references therein).

The electromagnetic fields of an arbitrary time dependent electric dipole (fixed at a given point in space, say, the origin) have many terms, not only the radiation term proportional to $1/r$. More specifically, the electric field of an electric dipole has three contributions, namely, the static zone contribution, proportional to $1/r^3$, the intermediate zone contribution, proportional to $1/r^2$ and the far zone or radiation contribution, proportional to $1/r$ (as we shall see, the corresponding magnetic field has only two contributions). Unfortunately, a general deduction of the exact electric and magnetic fields of an electric dipole is lacking from most undergraduate and graduate textbooks. \cite{Griffiths2,JacksonBook,SommerfeldBook, LandauLifshitzBook, PanofskyPhillipsBook, KonopinskyBook, FeynmanBook, ReitzMilfordCristyBook, SmytheBook, StrattonBook, MarionHealdBook} In some cases, the authors obtain the exact electromagnetic fields for the electric dipole by assuming a harmonic time dependence for the sources \cite{JacksonBook, StrattonBook, SmytheBook, KonopinskyBook} or by using a very particular model for the charge and current distributions. \cite{FeynmanBook, SommerfeldBook, MarionHealdBook} In other cases, the authors are interested only in the radiation fields. \cite{Griffiths2,LandauLifshitzBook, PanofskyPhillipsBook, ReitzMilfordCristyBook}
 As far as we know, a general deduction of the exact electric and magnetic fields of an electric dipole with arbitrary time dependence can be found in Heras' paper. \cite{HerasPaper4} In this paper, the author first presents a new version of Jefimenko's equations in a material medium \cite{JefimenkoPaper1} with polarization ${\bf P}$ and magnetization ${\bf M}$ and, then,  choosing appropriately the values of ${\bf P}$ and ${\bf M}$, obtains the exact dipole fields.

Our purpose in this note is to provide the reader with a very direct deduction of the exact electric and magnetic fields of an electric dipole with arbitrary time dependence and without making any particular assumption for the (localized) charge and current distributions of the system. Our starting point will be Jefimenko's equations and our procedure consists simply in making a multipole expansion up to the desired order. It is worth emphasizing that our procedure does not require Jefimenko's equation in matter, which makes our deduction extremely simple.

Jefimenko's equations for arbitrary but localized sources are given by \cite{JefimenkoBook, HerasPaper1, GriffithsHealdPaper, TranCongTonPaper, BornaticiBellottiPaper}
\begin{eqnarray}
{\bf E}({\bf r}, t)& = & \dfrac{1}{4\pi\epsilon_0}
                    \left\{\int d{\bf r'} \dfrac{[\rho({\bf r'}, t')]
                    {\bf R}}{R^3} + \int d{\bf r'} \dfrac{[\dot{\rho}({\bf r'}, t')]
                    {\bf R}}{cR^2} - \int d{\bf r'}
                    \dfrac{[\dot{{\bf J}}({\bf r'},t')]}{c^2R}\right\}\, ,
\label {electricfield}\\ \cr
{\bf B}({\bf r}, t)& = & \dfrac{\mu_0}{4\pi}\left\{\int d{\bf r'}
                    \dfrac{[{\bf J}({\bf r'}, t')] \times {\bf R}}{R^3}
                    + \int d{\bf r'} \dfrac{[\dot{{\bf J}}({\bf r'},t')] \times
                    {\bf R}}{cR^2}\right\}\, ,
\label{magneticfield}
\end{eqnarray}
where the integrals are over all space, $R = |{\bf R}| = |{\bf r}-{\bf r'}|$, the overdot means time derivative
 (that is, $\dot\rho({\bf r}^{\,\prime}, t^{\,\prime}) =  \mbox{\large$\frac{\partial\rho}{\partial t^{\,\prime}}$}
 ({\bf r}^{\,\prime}, t^{\,\prime})$ and so on) and the brackets $[...]$ mean that the quantities inside them
must be evaluated at the retarded time, namely, $t' =  t - R/c$.

Now, we shall make a multipole expansion which consists, essentially, in making an expansion in powers of $r^{\,\prime}/r$, which is valid outside the charge and current distributions. In this sense, our final result will be valid only outside the sources. However, since we are interested in the dipole fields, we shall retain only terms up to linear order in ${\bf r}^{\,\prime}$. Therefore, we may use the following Taylor expansions
\begin{eqnarray}
\dfrac{1}{R} &\simeq& \dfrac{1}{r} + \dfrac{(\hat{\bf r}
                \cdot{\bf r'})}{r^2} \label{expansion1}
\end{eqnarray}
\begin{eqnarray}
\dfrac{{\bf R}}{R^2} &\simeq& \dfrac{\hat{\bf r}}{r} +
            \dfrac{2(\hat{\bf r}\cdot{\bf r'})\hat{\bf r}}{r^2}
            - \dfrac{{\bf r'}}{r^2} \label{expansion2}
\end{eqnarray}
\begin{eqnarray}
\dfrac{{\bf R}}{R^3} &\simeq& \dfrac{\hat{\bf r}}{r^2} +
        \dfrac{3(\hat{\bf r}\cdot{\bf r'})\hat{\bf r}}{r^3} -
        \dfrac{{\bf r'}}{r^3}\label{expansion3}
\end{eqnarray}
\begin{eqnarray}
t - \frac{R}{c} &\simeq& t_0+\dfrac{\hat{\bf r}\cdot{\bf r'}}{c}
\label{expansao4}
\end{eqnarray}
where we defined $t_0 = t - r/c$, identified as the retarded time with respect to the origin. Analogously,
we expand the source terms around $t^\prime = t_0$ and keep terms only up to linear order in ${\bf r}^\prime$.
This leads to the following expressions
\begin{equation} \label{ExpansaoRho}
[\rho({\bf r'}, t')] = \rho({\bf r'}, t-R/c) \simeq \rho({\bf r'}, t_0 +
            \hat{\bf r}\cdot{\bf r'}/c) \simeq
            \rho({\bf r'},t_0) + \dfrac{\hat{\bf r}\cdot{\bf r'}}{c}\dot{\rho}({\bf r'},t_0)\, ,
\end{equation}
\begin{equation}
            [{\bf J}({\bf r'}, t')] = {\bf J}({\bf r'}, t-R/c) \simeq {\bf J}({\bf r'}, t_0 +
            \hat{\bf r}\cdot{\bf r'}/c) \simeq
            {\bf J}({\bf r'},t_0) + \dfrac{\hat{\bf r}\cdot{\bf r'}}{c}\dot{{\bf J}}({\bf r'},t_0)\, ,
\label{ExpansaoJ}
\end{equation}
where we used equation (\ref{expansao4}). We can make expansions completely analogous to the previous ones
 to obtain approximate expressions for the time derivatives of the sources, namely,
\begin{equation} \label{ExpansaoRhoPonto}
[\dot\rho({\bf r'}, t')]   \simeq
            \dot\rho({\bf r'},t_0) + \dfrac{\hat{\bf r}\cdot{\bf r'}}{c}\ddot{\rho}({\bf r'},t_0)\, ,
\end{equation}
\begin{equation}
            [\dot{\bf J}({\bf r'}, t')]  \simeq
            \dot{\bf J}({\bf r'},t_0) + \dfrac{\hat{\bf r}\cdot{\bf r'}}{c}\ddot{{\bf J}}({\bf r'},t_0)\, ,
\label{ExpansaoJPonto}
\end{equation}
For convenience, let us denote the three integrals that appear on the right hand side (rhs) of equation (\ref{electricfield}) by $I_1^{(E)}$, $I_2^{(E)}$ and  $I_3^{(E)}$, respectively. In a similar way, let us denote by  $I_1^{(B)}$ and $I_2^{(B)}$, respectively, the two integrals that appear on the rhs of equation (\ref{magneticfield}). We must, then, compute these five integrals up to the desired order. Let us start with $I_1^{(E)}$. Substituting equations (\ref{expansion3}) and (\ref{ExpansaoRho}) into the expression for  $I_1^{(E)}$, we get
\begin{eqnarray}
I_1^{(E)} &\simeq& \int d{\bf r'} \left\{ \rho({\bf r'},t_0) +
                                        \dfrac{\hat{\bf r}\cdot{\bf r'}}{c}\dot{\rho}({\bf r'},t_0)\right\}
                                        \left\{ \dfrac{\hat{\bf r}}{r^2} + \dfrac{3(\hat{\bf r}\cdot{\bf r'})
                                        \hat{\bf r}}{r^3} - \dfrac{{\bf r'}}{r^3}\right\} \cr\cr
&\simeq&\dfrac{\hat{\bf r}}{r^2}\left\{\int d{\bf r'} \rho({\bf r'},t_0)\right\}
                                        + \dfrac{3\hat{\bf r}}{r^3}\left\{\hat{\bf r}\cdot\int d{\bf r'}
                                        \rho({\bf r'},t_0){\bf r'}\right\}\cr \cr
                                        &-& \dfrac{1}{r^3}\left\{\int d{\bf r'}
                                        \rho({\bf r'},t_0){\bf r'}\right\}
                                        + \dfrac{\hat{\bf r}}{cr^2}\left\{\hat{\bf r}\cdot\int d{\bf r'}
                                        \dot{\rho}({\bf r'},t_0){\bf r'}\right\}\, ,
\label{expansion6}
\end{eqnarray}
where quadratic terms in ${\bf r'}$ were neglected. Note that the integral in the first term on the rhs of last equation is simply the total charge of the distribution which is independent of time due to charge conservation. This term corresponds to the monopole term of the multipole expansion, since it is the field created by a point charge  $Q:=\int\rho({\bf r}^\prime, t_0) \, d{\bf r}^\prime$ fixed at the origin. Of course, this term does not contribute to the dipole field. In the next two integrals on the rhs of last equation we identify immediately the electric dipole moment of the distribution at instant $t_0$, namely, ${\bf p}(t_0) =  \int d{\bf r}^\prime\, \rho({\bf r}^\prime, t_0) \, {\bf r} ^\prime $.
 Finally, in the last integral on the rhs of last equation, after we commute the time derivative with the integration, we identify the time derivative of the electric dipole moment of the distribution at instant $t_0$, namely, $\dot{\bf p}(t_0) = \mbox{\large$\frac{d}{dt}$} \int d{\bf r}^\prime\, \rho({\bf r}^\prime, t_0) \, {\bf r} ^\prime $. Hence, integral $I_1^{(E)}$ is given by
\begin{equation}
I_1^{(E)} \simeq \dfrac{Q\hat{\bf r}}{r^2}+\dfrac{3[\hat{\bf r}
                    \cdot {\bf p}(t_0)]\hat{\bf r}-{\bf p}(t_0)}{r^3}+
                     \dfrac{[\hat{\bf r}\cdot \dot{\bf {p}}(t_0)]}{cr^2}\, .
\label{integral1}
\end{equation}
Using a procedure  analogous to the previous one, but now substituting into the expression of $I_2^{(E)}$ the approximations written in (\ref{expansion2}) and (\ref{ExpansaoRhoPonto}), it is straightforward to show that
\begin{equation}
I_2^{(E)} \simeq \dfrac{2[\hat{\bf r}\cdot \dot{\bf {p}}(t_0)]
                                        \hat{\bf r}-\dot{\bf {p}}(t_0)}{cr^2}+
                                        \dfrac{[\hat{\bf r}\cdot \ddot{\bf {p}}(t_0)]
                                        \hat{{\bf r}}}{c^2r}\, .
\label{integral2}
\end{equation}

In order to compute the remaining three integrals, $I_3^{(E)}$, $I_1^{(B)}$ e $I_2^{(B)}$, it will be convenient to write the cartesian basis vectors as $\hat{{\bf e}}_i = {\bf \nabla}'{x'_i}$,  $(i=1,2,3)$, so that any vector ${\bf b}$ can be written in the form ${\bf b} = b_i\hat{{\bf e}}_i = ({\bf b} \cdot \hat{{\bf e}}_i)\,\hat{{\bf e}}_i = ({\bf b} \cdot {\bf \nabla}'{x'_i})\,\hat{{\bf e}}_i$, where Einstein convention of summation over repeated indices is assumed. As it will become clear in a moment, in order to keep only linear terms in ${\bf r'}$, when computing $I_3^{(E)}$ it suffices to make in the integrand the approximations
 ${R} \rightarrow {r}$ and $[\dot{{\bf J}}({\bf r'},t^\prime)] \rightarrow \dot{{\bf J}}({\bf r'},t_0)$. Hence, we have
\begin{eqnarray}\label{IE3}
I_3^{(E)}& \simeq& -\dfrac{1}{c^2r}\int d{\bf r'} \dot{\bf J}({\bf r'},t_0)\cr
                &=&
                -\dfrac{1}{c^2r} \ \hat{\bf e}_i\int d{\bf r'} \{\dot{\bf J}({\bf r'},t_0)
                \cdot \nabla'x'_i\} \cr
                &=& \dfrac{1}{c^2r} \ \hat{\bf e}_i\int d{\bf r'}  x'_i \{\nabla'\cdot\dot{\bf J}
                ({\bf r'},t_0)\}\cr
                &=&
                 -\dfrac{1}{c^2r} \ \hat{\bf e}_i\int d{\bf r'}  x'_i \ddot{\rho}
                ({\bf r'}, t_0)
\end{eqnarray}
where in passing from the second line to the third one we made an integration by parts and discarded the surface term, since the integration is
over all space and the sources are localized. In the last step we used the continuity equation. Identifying in (\ref{IE3}) the second time derivative of the electric dipole moment of the distribuition at instant $t_0$, we obtain
\begin{eqnarray}
I_3^{(E)} \simeq -\dfrac{\ddot{{\bf p}}(t_0)}{c^2r}
\label{integral3}
\end{eqnarray}
Now we see why it was not necessary to go beyond zeroth order in the expansion of $[\dot{{\bf J}}({\bf r'},t^\prime)]$ in the computation of
 $I_3^{(E)}$: the electric dipole moment of the distribution is already of linear order in ${\bf r'}$, so that the next order term would contribute to the next terms of the multipole expansion, namely, the magnetic dipole term and the electric quadrupole term.

 In the computation of the last two integrals, $I_1^{(B)}$ and $I_2^{(B)}$, which will give us the contribution to the magnetic field of the electric dipole term, we must proceed exactly as we did in the case of  $I_3^{(E)}$, namely, it suffices to use only the zeroth order approximation for the expansions of the current and its time derivative. Hence, we have for $I_1^{(B)}$,
\begin{eqnarray}
I_1^{(B)} &\simeq& -\dfrac{\hat{{\bf r}}}{r^2} \times \left\{\int d{\bf r'} \
                    {\bf J}({\bf r'}, t_0)\right\}\cr
                    &=&
                     -\dfrac{\hat{{\bf r}}}{r^2}
                    \times \ \hat{\bf e}_i\int d{\bf r'} \{{\bf J}({\bf r'},t_0) \cdot \nabla'x'_i\} \cr
                    &=&
                    \dfrac{\hat{{\bf r}}}{r^2} \times \ \hat{\bf e}_i\int d{\bf r'}  x'_i
                    \{\nabla'\cdot{\bf J}({\bf r'},t_0)\} \cr
                    &=&
                     -\dfrac{\hat{{\bf r}}}{r^2} \times \
                    \hat{\bf e}_i\int d{\bf r'}  x'_i \dot{\rho}({\bf r'}, t_0)\cr
                    &=&
                    \dfrac{\dot{\bf {p}}(t_0) \times \hat{{\bf r}}}{r^2}
\label{integral4}
\end{eqnarray}
where in the last step we identified  the time derivative of the electric dipole moment of the distribuition.
Proceeding in a completely analogous way, it is straightforward to show that
\begin{equation}
I_2^{(B)} \simeq \dfrac{\ddot{\bf {p}}(t_0) \times \hat{{\bf r}}}{cr}\, .
\label{integral5}
\end{equation}

Inserting results (\ref{integral1}), (\ref{integral2}) and (\ref{integral3}) into equation (\ref{electricfield}), discarding the monopole term, as well as inserting results (\ref{integral4}) and (\ref{integral5}) into equation (\ref{magneticfield}), we finally obtain the exact electric and magnetic fields associated to the electric dipole term of an arbitrary (localized) distribution of charges and currents, namely,
\begin{eqnarray}
{\bf E}({\bf r},t)\!\!\!&=&\!\!\!\dfrac{1}{4\pi\epsilon_0}\left\{
                    \dfrac{3[\hat{\bf r}\cdot {\bf p}(t_0)]\hat{\bf r}-{\bf p}(t_0)}{r^3}+
                    \dfrac{3[\hat{\bf r}\cdot \dot{\bf {p}}(t_0)]\hat{{\bf r}}-\dot{\bf
                    {p}}(t_0)}{cr^2} + \dfrac{\hat{\bf r} \times [\hat{{\bf r}}
                    \times\ddot{\bf {p}}(t_0)]}{c^2r}\right\}\, ,
\label{CampoEFinal} \\ \cr
                    {\bf B}({\bf r},t)\!\!\!&=&\!\!\!\dfrac{\mu_0}{4\pi}
                    \left\{ \dfrac{\dot{\bf {p}}(t_0) \times \hat{{\bf r}}}{r^2} +
                    \dfrac{\ddot{\bf {p}}(t_0) \times \hat{{\bf r}}}{cr} \right\}\, ,
\label{CampoBFinal}
\end{eqnarray}
which are the desired fields (expression (\ref{CampoEFinal}) is written, but not demonstrated, in Milonni's book,\cite{MilonniBook} for instance).
A few comments are in order here.

 Firstly, it is worth mentioning that the last terms on the rhs of equations (\ref{CampoEFinal})
and (\ref{CampoBFinal}), those which are proportional to $1/r$, correspond to the radiation fields of an electric dipole.
Secondly, for the idealized case of a point electric dipole fixed at the origin equations (\ref{CampoEFinal})
and (\ref{CampoBFinal}) are exact for every point of the space, except the origin, where the fields are singular. However, since we are always interested in the fields outside the dipole, we shall not be concerned with these singular terms (a discussion of this kind of term for the case of a static point electric dipole can be found in chapter 3 of Jackson's textbook \cite{JacksonBook}).
Finally, as an important particular case, we consider an electric dipole with harmonic time dependence.
Substituting ${\bf p}(t) = {\bf p_0}e^{-i\omega t}$ into the previous equations,
we obtain the well known result \cite{JacksonBook}
\begin{eqnarray}\label{CampoEFinalHarmonico}
{\bf E}({\bf r},t)\!\!\!&=&\!\!\!\dfrac{1}{4\pi\epsilon_0}
								\left\{ k^2(\hat{{\bf r}}\times {\bf p})\times
								\hat{{\bf r}} \dfrac{e^{ikr}}{r} +
								[3(\hat{{\bf r}}\cdot {\bf p})\hat{{\bf r}}-
								{\bf p}]\left(\dfrac{1}{r^3}-\dfrac{ik}{r^2}\right)e^{ikr}\right\}\, ,
								\\ \cr
{\bf B}({\bf r},t)\!\!\!&=&\!\!\!\dfrac{\mu_0}{4\pi}\left\{
								ck^2(\hat{{\bf r}}\times{\bf p})\dfrac{e^{ikr}}{r}
								\left(1-\dfrac{1}{ikr}\right) \right\}		\, ,\label{CampoBFinalHarmonico}						
\end{eqnarray}
where $k=\omega/c$. Note the presence in equation (\ref{CampoEFinal}) of a term proportional to $r^{-3}$ and ${\bf p}(t_0)$, which is characteristic of the field of a static electric dipole, except by the fact that here the electric dipole moment is evaluated at retarded time $t_0$. This term dominates in the near zone, where $d\ll r \ll \lambda$, being $d$ a typical length scale of the source and $\lambda$ the wavelength of the electromagnetic field. Observe the absence of such a term in equation (\ref{CampoBFinal}), which could have been anticipated, since a static electric dipole does not create any magnetic field. In the radiation zone, where $d\ll \lambda \ll r$, the dominating terms are proportional to
 $r^{-1}$ and to the second time derivative of the electric dipole moment. Further, it is evident the transverse character of the radiation fields.

The results given by equations (\ref{CampoEFinal}) and (\ref{CampoBFinal}) could have been obtained if, instead of the Jefimenko's equation (\ref{electricfield}) for the electric field,  we had used the equivalent
Panofsky-Phillips equation for the electric field, \cite{PanofskyPhillipsBook} given by
\begin{eqnarray}
{\bf E}({\bf r},t) &=& \dfrac{1}{4\pi \epsilon_0}\left\{\int d{\bf r'}
                                        \dfrac{[\rho({\bf r'},t')] \hat{{\bf R}}}{R^2}
                                     + \int d{\bf r'} \dfrac{([{\bf J}({\bf r'},t')]
                                     \cdot \hat{{\bf R}})\hat{{\bf R}} + ([{\bf J}({\bf r'}, t')]
                                     \times \hat{{\bf R}})\times \hat{{\bf R}}}{cR^2}\right\} \cr \cr
                              &+& \dfrac{1}{4\pi \epsilon_0}\left\{\int d{\bf r'} \dfrac{([\dot{{\bf J}}({\bf r'},t')]
                              \times \hat{{\bf R}})\times \hat{{\bf R}}}{c^2R}\right\}\, .
\label{PanofskyPhillipsequation}
\end{eqnarray}
It can be shown that the above equation is indeed equivalent to Jefimenko equation (\ref{electricfield}) for the electric field
 \cite{McDonaldPaper} and it is more convenient as a starting point for computing multipole radiation fields.  \cite{ReinaldoPaper}
  In fact, all multipole radiation fields can be obtained from the  last terms of the rhs of equations (\ref{PanofskyPhillipsequation}) and
  (\ref{magneticfield}). The dipole radiation fields, identified previously in equations (\ref{CampoEFinal}) and (\ref{CampoBFinal}), are just the first contributions of the whole multipole expansion for the radiation fields. The next order contributions, namely, the radiation fields of a magnetic dipole and an electric quadrupole may be similarly obtained with no much more effort. \cite{ReinaldoPaper}

  We leave, for the interested reader, the exercise of reobtaining the exact electric  field of an electric dipole with arbitrary time dependence from equation (\ref{PanofskyPhillipsequation}) instead (\ref{electricfield}). The exact electric and magnetic fields of higher multipoles, as for instance, magnetic dipole and electric quadrupole, can also be obtained following our approach. Of course, for these cases, when making the necessary Taylor expansions, higher order terms in ${\bf r}^\prime$ must be kept.

In this note we have computed the exact electric and magnetic fields of an electric dipole with arbitrary time dependence, in contrast with the usual calculations found in undergraduate and graduate textbooks where a harmonic time dependence is assumed. It is worth mentioning that most textbooks base their discussions in the electromagnetic potentials. Here, instead, we have made a multipole expansion directly from Jefimenko's equations. Besides the simplicity of our method, we think it enlarges the list of problems that can be attacked directly by using Jefimenko's equations (see also the papers \cite{HerasPaper2,HerasPaper3,HerasPaper5,RohrlichPaper1,RohrlichPaper2,JefimenkoPaper}), avoiding this way the introduction of  electromagnetic potentials.

\begin{acknowledgments}
We are indebted to Prof. C. Sigaud for reading the manuscript. We also would like to thank CNPq for partial financial support.
\end{acknowledgments}

\end{document}